\documentclass[runningheads, a4paper, 10pt]{llncs}

\usepackage{amssymb}
\usepackage{graphicx}
\usepackage{subfigure}
\usepackage{caption}
\usepackage{hyperref}
\usepackage{paralist}
\usepackage{color}

\usepackage{amsmath}
\date{}
\renewcommand{\baselinestretch}{1.1}

\newcommand{\Prol}{\ensuremath{\mathsf{Prol}}} 
\newcommand{\Aver}{\ensuremath{\mathsf{Mod}}} 
\newcommand{\Low}{\ensuremath{\mathsf{Low}}}
\newcommand{\Dual}{\ensuremath{\mathsf{Dual}}}
\begin{document}

\author{Mishari Almishari$^\ddag$, Mohamed Ali Kaafar$^{\ddag\ddag}$, Gene Tsudik$^\dag$,  Ekin Oguz$^\dag$}
\institute{$^\ddag$ King Saud University\\$^{\ddag\ddag}$NICTA\\$^\dag$University of California, Irvine}

\title{Are 140 Characters Enough? \\ A Large-Scale Linkability Study of Tweets}
\maketitle

\begin{abstract}
Microblogging is a very popular Internet activity that informs and entertains great multitudes
of people world-wide via quickly and scalably disseminated terse messages containing
all kinds of news-worthy utterances. Even though microblogging is neither designed nor meant
to emphasize privacy, numerous contributors hide behind pseudonyms and compartmentalize
their different incarnations via multiple accounts within the same, or across multiple, site(s).

Prior work has shown that stylometric analysis is a very powerful tool capable of linking product 
or service reviews and blogs that are produced by the same author when the number of authors is large. 
In this paper, we explore linkability of tweets.
Our results, based on a very large corpus of tweets, clearly demonstrate that, at least
for relatively active tweeters, linkability of tweets by the same author is easily attained 
even when the number of tweeters is large. We also show that our 
linkability results hold for a set of {\em actual Twitter users} who tweet from multiple accounts.
This has some obvious privacy implications, both positive and negative.
\end{abstract}
\section{Introduction}
\label{intro}
%


Microblogs offer a fast and highly scalable information sharing, allowing multitudes of users to 
disseminate pithy messages to news-hungry and attention-challenged followers or subscribers. 
For either social or professional purposes, users share their thoughts, interests and sometimes
express highly sensitive or controversial opinions. Twitter, the most prominent microblogging site, 
has now grown to a truly global service with hundreds of millions of users \cite{twitterusers}. 
In Twitter, Weibo Sina and others, the relationship between users (referred to as tweeters in Twitter) 
typically requires no reciprocal approval and is often considered as a form of subscription. 
Although postings (called tweets in Twitter) can be tagged as private, most users keep 
theirs public, as well as their follower information. 

At the same time, microblogging has become a rich source of information about individuals. 
Although it may be difficult to justify the claim that the existence of public messages 
violates privacy of their authors (since they are the one who make their utterances public in 
the first place), the potential to de-anonymize multiple user accounts and to link messages 
(or sets thereof) produced by the same author represents a threat to privacy. 
This is primarily because multiple accounts owners often expect each set of messages 
to remain within the boundaries of the account from which it has been posted.  

Consider for example, the case of an activist who uses a pseudonym in Twitter\footnote{Twitter, as 
opposed to other major OSNs, does not require users to provide real names; pseudonyms are 
accepted as long as they do not impersonate other users accounts \url{https://support.twitter.com/entries/18311}}, 
to post some highly sensitive or controversial tweets and, in parallel, uses another account associated 
with his/her real name. Linkage of these two accounts might pose a serious threat to that 
person's privacy and perhaps even to their physical security.    

On the other hand, microblogging site operators and law-enforcement agencies might benefit 
from techniques that link accounts within a site or across multiple sites. Legitimate reasons might 
include: (1) identifying spammers and phishers who might hide behind multiple accounts 
to evade detection, (2) tracking or correlating messages pertaining to illegal activities\footnote{Clearly, 
the definition of illegal activities varies widely from country to country.}, such as terrorism incitement, 
pedophilia or human or drug trafficking.  

In this paper, we investigate the use of probabilistic models to link user accounts in the particular 
microblogging ecosystem of Twitter, where message size is limited to
140 characters and where both the use of metadata tags (hashtags) and reposts of 
other users' messages (retweeting) is commonplace.  


The goal of this work is to assess linkability of tweets by measuring how much they relate/link to author's 
other tweets. We  perform  our analysis over two large datasets, each containing over $8,000$ Twitter 
accounts, with over $28$ million tweets in one set and more than $3$ million in the other. 
The first dataset consists of prolific tweeters, with users producing $2,000$ tweets or more over a 
six-months period. The second dataset consists of less prolific tweeters who post between 
$300$ and $400$ tweets, over the same period. Furthermore, we extend our analysis to show the 
linkability of tweets of actual users who operate several Twitter accounts.  
Our work makes the following contributions:
%
\begin{enumerate}
\item We conduct a large-scale linkability study by applying the Na\"ive Bayes model as a classifier to predict linkage between tweets using 
simple stylometric features and analyzing the effect of a specific Twitter feature -- hashtags.
Results demonstrate that tweets are highly linkable. By simply analyzing letter frequencies, 
our classifier can -- in some scenarios -- link tweets at a large scale with over
$90\%$ accuracy. We also show that hashtags are particularly useful in the linkage process: 
by only using hashtags, our classifier achieves high linkability ratios and can link two thirds of the tweets.       
\item We extend and verify our linkability technique to link tweets of dual Twitter accounts owners. We consider a set of tweets from 14 Twitter users who maintain and simultaneously use two accounts or more, and successfully classify and extract linkage of their tweets amongst sets of over 8000 users.
\end{enumerate}
We envision several possible uses of our technique. First, it could be implemented as a 
service usable by prospective tweeters to assess linkability of their multiple accounts. 
Second, it can be used by the provider (i.e., Twitter) to identify multiple accounts 
holders who generate libel, spread disinformation or promote illegal activities.  

\section{Background: Na\"ive Bayes Model}
\label{background}
This section overviews the Na\"ive 
Bayes Model \cite{nbayes} used in the subsequent linkability analysis.

%
Na\"ive Bayes (NB) is a probabilistic model based on the so-called Na\"ive Bayes assumption,  
which states that all features/tokens are conditionally independent, given some category. 
In our case, the category is a tweeter (user). Given a document with a set of tokens/features:  
$token_1, token_2,\ldots,token_n$, NB model computes the 
corresponding user model as follows:
\begin{equation}
\footnotesize
User = arg max_U P(U|token_1,\ldots,token_n) \nonumber
\label{eq:nb}
\end{equation}
where $U$ varies over all distinct users in our dataset, in order to maximize the 
probability $P(U|token_1,\ldots,token_n)$ of categorizing a document
(represented by $token_1, token_2,\ldots,token_n$) as belonging to a specific user. 
Using the Bayes Rule\cite{mlearn1}, $P(U|token_1,\ldots,token_n)$ is defined as:
\footnotesize
\begin{eqnarray} \label{frac}
P(U|token_1,token_2,\ldots,token_n)= \frac{P(U)P(token_1,token_2,\ldots,token_n|U)}{P(token_1,\ldots,token_n)} \nonumber
\end{eqnarray}
\normalsize
where $P(token_1,token_2,\ldots,token_n|U)$ is the probability of generating the set 
of tokens $token_1,\ldots,token_n$ by $U$. When using the Na\"ive Bayes assumption,
\footnotesize
$P(token_1,\ldots,token_n|U) = \nonumber P(token_1|U)\cdots P(token_n|U) \nonumber,$
\normalsize finding a matching $User$ for $token_1,$ $token_2$, $\ldots,token_n$ 
for a given tweeter profile boils down to:
\begin{eqnarray}
\footnotesize
&User = argmax_U P(token_1|U)\cdots P(token_n|U) \nonumber
\label{eq:symkl}
\end{eqnarray}
\normalsize
where $P(token_i | U)$ is the probability of generating $token_i$ by $U$.
We assume that $P(U)$ -- the probability of associating
a document to a user without any information about the tokens in the document --
is the same for all tweeters. Also,
the denominator $P(token_1,\ldots,token_n)$ in the equation above is the combined 
probability of all tokens; it is therefore independent of the user. 
Finally, to avoid the under-flow problem, we consider the 
$log$ of the products, which results in:
\begin{eqnarray} 
\footnotesize
&User = argmax_U \sum_i logP(token_i|U) \nonumber 
\end{eqnarray}
\normalsize
We estimate all probabilities of the form $P(token_i|U)$ using the Maximum Likelihood 
estimator \cite{bishopbook} with Laplace smoothing \cite{mlearn1}.

\section{Dataset}
\label{dataset}
We use a portion of the dataset crawled by Yang et al. \cite{leskovec-wsdm-2011}, that
spans an approximately six-months period from June to December, 2009, 
excluding October\footnote{Due to some  technical difficulties in extracting data for October 2009}.

\begin{table}[th!]
\begin{center}
\begin{tabular}{|r|l|}
\hline
{\bf Parameter} & {\bf Value}
\\ \hline
Total \# of Tweets  & $400,834,808$ 
\\ \hline
Total \# of Tweeters& $15,905,473$
\\ \hline
Max. \# of Tweets per Tweeter & $82,177$
\\ \hline
Min. \# of Tweets per Tweeter & $1.0$
\\ \hline
\%  Tweeters with $\geq 2000$ Tweets & $0.05\%$
\\ \hline
\% Tweeters with $\geq  500$  Tweets & $0.68\%$
\\ \hline
\% Tweeters with $\geq  300$  Tweets & $1.4\%$
\\ \hline
\% Tweeters with $\geq 50$ Tweets & $9.5\%$
\\ \hline
\% Tweeters with $\leq 10$ Tweets & $73\%$
\\ \hline
\% Tweeters with $\leq 1$ Tweets & $33\%$
\\ \hline
\end{tabular}
\end{center}
\caption{Dataset Statistics}
\label{tab:stat}
\end{table}  

The dataset consists of more than $400$ million tweets, authored by over $15$ million users 
(see Table \ref{tab:stat}). Majority of tweeters authored no more than $10$ tweets each, 
i.e., most are not what we would call prolific tweeters. However, there is still a substantial 
number of prolific tweeters. 

For analysis purposes, we extracted two subsets. The first is referred to as $\Prol$. It contains 
all tweets of all users who authored at least $2,000$ tweets during the observed 6-month period. 
This set consists of $8,262$ different twitter accounts. We consider these to be highly-prolific tweeters. 
The total number of tweets in $\Prol$ is $28,625,352$.

The second subset is referred to as $\Low$. It contains all tweets authored by a set of 
$10,000$ users, randomly chosen among all users who authored between $300$ and $400$ 
tweets. This corresponds to a total of $73,004$ users -- much broader demographic. 
The total number of tweets in $\Low$ is $3,449,635$. Since the original dataset is from 
2009, given ever-increasing popularity of Twitter, we speculate that $\Low$ is a sample of 
very large demographic in the current state of Twitter. 

We extracted two subsets, instead of one, since we aim to assess linkability of tweets for 
users in different prolificacy scales. We acknowledge that our selection of thresholds in 
constituting these subsets is subjective. However, we believe that increasing use of Twitter 
\cite{twitterusers} results in more users falling into the ranges of these thresholds. 
Linkability analysis for users who produce fewer tweets is deferred to future work. 

\section{Settings and Methodology}
\label{sec:settings}
We adopt the settings, linkability-related definitions and abbreviations similar to those
in \cite{mish12}. Our goal is to first assess how much tweets say about their authors, i.e., 
how accurately a seemingly anonymous recent tweet can be linked to previous tweets 
by the same author. Later, in Section \ref{realworld}, we analyze how tweets authored 
by the same tweeter, while using two different accounts, can be re-linked. 

We  build a Na\"ive Bayes classifier as our matching/linking model, which is partially trained on 
two subsets: $\Prol$ and $\Low$, to perform the linkage. Specifically, for each author $U$ in 
both $\Prol$ and $\Low$, we randomly split her tweets into two sets: Identified Record (IR), and 
Anonymous Record (AR). The set of all IRs is used for training the classifier and the set of all 
ARs is used to assess linkability -- the accuracy of linking an AR to its corresponding IR. 
Note that we conduct linkability analysis independently over each set.  
 
For each set and for each author, the task of the classifier is to link her AR to a corresponding 
IR while maximizing the number of ARs correctly linked. For each set, it matches each author's 
AR to an IR by returning a list of candidate IRs, sorted in decreasing order of likelihood of being 
the correct match. 

We consider a given AR to have Top-$x$ linkability if the actual corresponding IR is among 
the top $x$ candidate IR records that the classifier returns. We measure performance of 
our classifier in terms of linkability ratio (LR), which computes the percentage of ARs 
that have been correctly classified within the top $x$ candidates. 
We experimented with three $x$ values:  $1$, $5$, and $10$. 

We partitioned each author's tweets into IR and AR as follows: First, we sorted a user's tweets 
in random order. Then, we assigned all tweets (except the last $100$) to IR. For the last $100$ 
tweets, we assigned the first $y$ to AR. We vary $y$ over the following values: 
$5$, $10$, $20$, $50$ and $100$.  The main reason for varying $y$ is to evaluate the 
impact of the number of anonymous tweets on linkability. 

\section{Linkability Analysis}
\label{analysis}
We use the Na\"ive Bayes (NB) model as a matching tool to link tweets based on two types of lexical tokens: (i) unigrams: all letters of the English alphabet, i.e., $26$ tokens. (ii) bigrams: all possible two-letter combinations, 
i.e., $676$ tokens. We perform separate analysis on $\Prol$ and $\Low$, and compare respective results. 
For each set, we build a NB model based on all IRs of the set, match all ARs and compute the resulting LR. 

\subsection{Unigrams}
\label{sec:uni}
\begin{figure}[thb!]
  \centering
  \subfigure[]{\label{fig:ng-1-prol}\includegraphics[height=2.0in,width=0.49\textwidth]{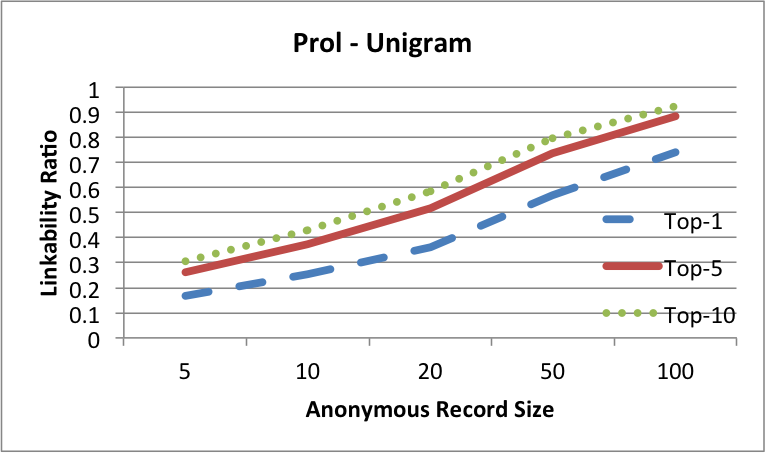}}
  \subfigure[]{\label{fig:ng-1-low}\includegraphics[height=2.0in,width=0.49\textwidth]{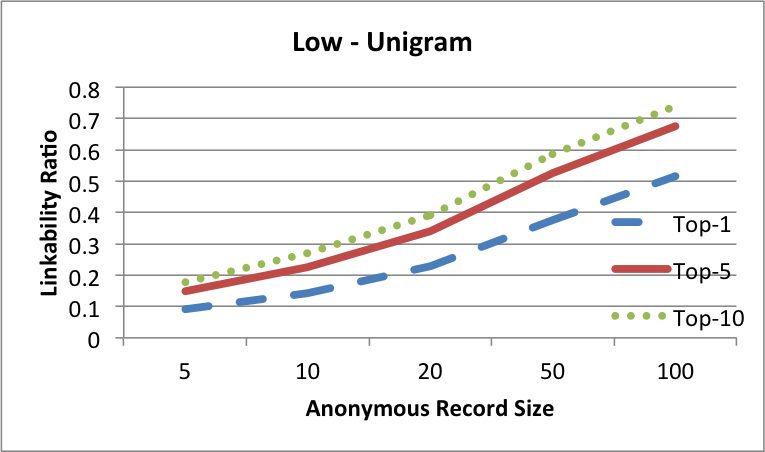}}	  
\label{fig:uni}
  \caption{Top-1, Top-5, and Top-10 LRs of unigram-based NB model for $\Prol$ \subref{fig:ng-1-prol}  and  $\Low$ \subref{fig:ng-1-low} }
\end{figure}

Figures \ref{fig:ng-1-prol} and \ref{fig:ng-1-low} show LRs for $\Prol$ and $\Low$, respectively. 
Specifically, they show Top-1, Top-5 and Top-10 LRs for various AR sizes, using unigrams in NB.  

First, we observe that all curves exhibit clear upward trend, showing that the bigger the AR size, 
the higher LR we obtain. For AR size of $100$, Top-10 LR is as high as 92\% for $\Prol$ and 74\% for $\Low$.  
Additionally, we observe relatively high LRs for Top-1(Top-5) -- 74\% (88\%) of the ARs  are linked in $\Prol$ 
for AR size of 100. Even for small AR sizes, we link a large number of ARs. For example, for AR size of 
$20$, Top-10 LR is $59\%$ for $\Prol$. Even though the number of tweeters is large and the number of 
tokens is only 26, we can link a substantial number of ARs. As expected, we observe that LRs 
are higher in sets that have larger IRs ($\Prol$ better than $\Low$), since larger IRs offer the classifier more 
information to capture the user's writing style.
 
\subsection{Bigrams}
\label{sec:di}
\begin{figure}[thb!]
  \centering
  \subfigure[]{\label{fig:ng-2-prol}\includegraphics[height=2.0in,width=0.49\textwidth]{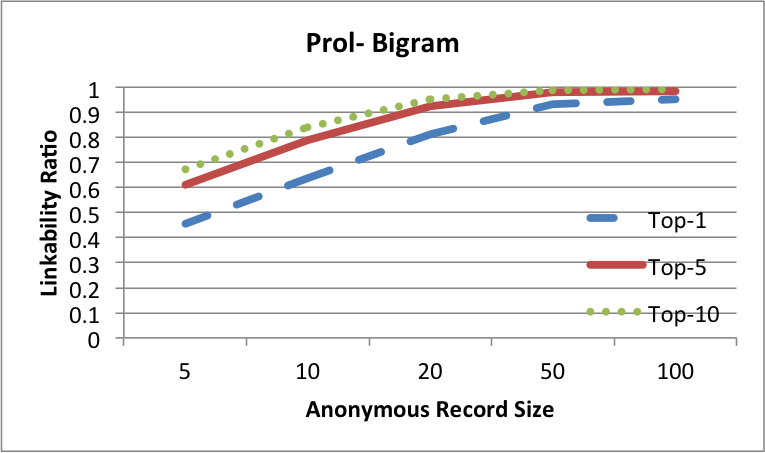}}
  \subfigure[]{\label{fig:ng-2-low}\includegraphics[height=2.0in,width=0.49\textwidth]{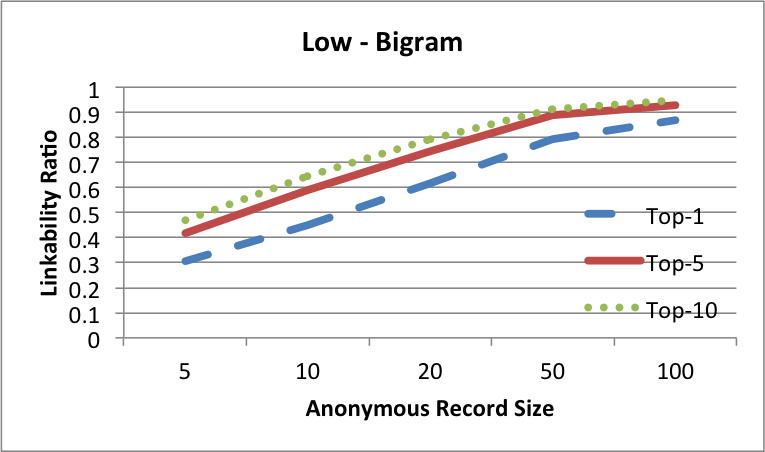}}
   \label{fig:bi}
  \caption{Top-1, Top-5, and Top-10 LRs of bigram-based NB model for $\Prol$ \subref{fig:ng-2-prol} and $\Low$ \subref{fig:ng-2-low}}
\end{figure}

Figures \ref{fig:ng-2-prol}  and \ref{fig:ng-2-low} show Top-1, Top-5 and Top-10 LRs, 
when bigrams are used. Substituting unigrams with bigrams substantially 
increases LRs, even when using a rather small AR.
For instance, Top-1 LR is 95\% and 87\% for $\Prol$ and $\Low$, respectively, for AR size of 100. 
Even for small AR sizes, we achieve high LRs. For example, for AR size of 5(10), Top-10 LR 
exceeds 67(84)\% in $\Prol$. Also, for AR size of only 20, Top-5 (Top-10) LR in $\Low$ is around 75(80)\%. 
As observed earlier with unigrams, sets with larger IRs offer higher LRs, which is again due to offering more 
information to model user's tweets. Notably, when using bigrams, the classifier needs a smaller AR size to achieve substantially higher LRs (e.g. using bigrams with an AR size of 20, the classifier obtains a Top-5 LR of over 70\% for $\Low$ tweeters, while it needs an AR of 100 or more to obtain similar performance with unigrams). 
These results suggest that there is a compromise in choosing between unigrams and bigrams. On one hand, bigrams lead to better linkability ratios with a less number of tweets to learn from. On the other hand, unigrams are less computationally demanding and should be considered when there is a need for a higher scalability and faster computation.   

\subsection{Varying the Number of Users}
\label{sec:vary}
In the previous sections, we  considered the full set of users, i.e., we included all $8,262$ users in 
$\Prol$ and $10,000$ --  in $\Low$. We now reduce the number of users in the sets and 
explore the effect of user set size on linkability.  

\noindent

\begin{figure}[thb!]
  \centering
  \subfigure[]{\label{fig:ng-1-vary-prol}\includegraphics[height=2.0in,width=0.49\textwidth]{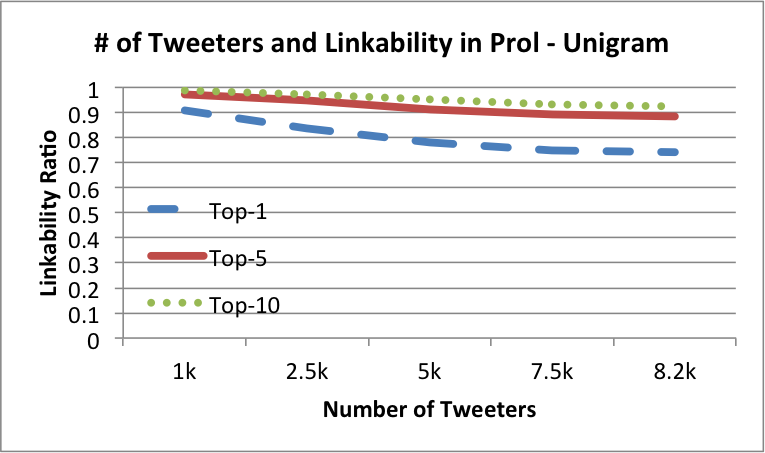}}
  \subfigure[]{\label{fig:ng-1-vary-low}\includegraphics[height=2.0in,width=0.49\textwidth]{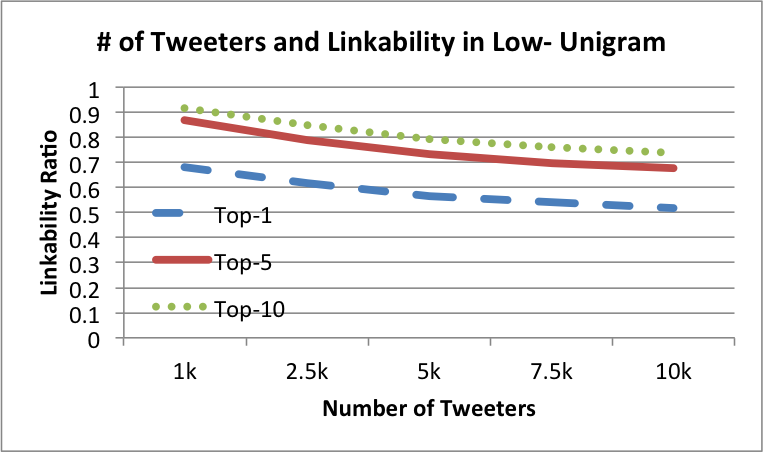}}
  \label{fig:uni-vary}
  \caption{ LRs of unigram-based NB model  when varying the number of users in  $\Prol$ \subref{fig:ng-1-vary-prol} and $\Low$ \subref{fig:ng-1-vary-low}}
\end{figure}

{\bf Unigrams.} We vary the number of users in $\Prol$ and $\Low$ between $1,000$ users and full set size (i.e. $8,262$ and $10,000$ users respectively).
Figures \ref{fig:ng-1-vary-prol} and \ref{fig:ng-1-vary-low} show LRs for different set sizes when AR size is $100$. 
As expected, LR increases with the smaller number of users. Top-10 LR reaches 99\% and 92\% in 
$\Prol$ and $\Low$, respectively, when the set size is $1,000$, which is reasonably large. When the set size is 
$5,000$, Top-5 LR exceeds 90\% and 70\% in $\Prol$ and $\Low$, respectively. Note that when 
we increase to the full set size, reduction of Top-10 LR does not exceed  
7\% and 20\% in $\Prol$ and $\Low$, respectively. This points to the resilience of our linkability model that is based on unigrams.  

\begin{figure}[thb!]
  \centering
  \subfigure[]{\label{fig:ng-2-vary-prol}\includegraphics[height=2.0in,width=0.49\textwidth]{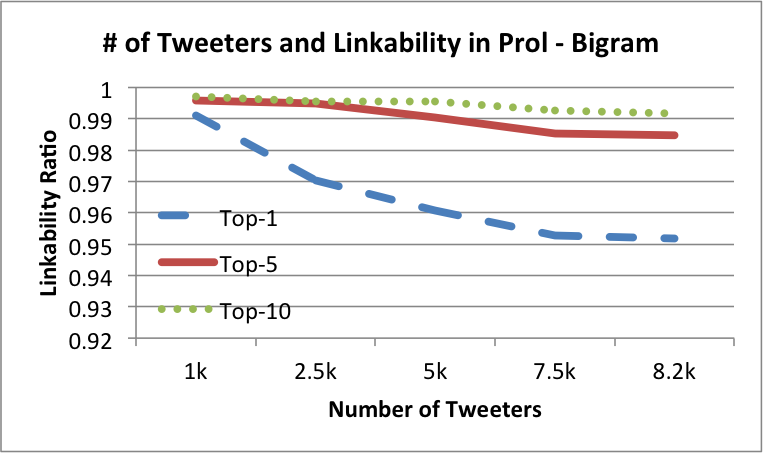}}
  \subfigure[]{\label{fig:ng-2-vary-low}\includegraphics[height=2.0in,width=0.49\textwidth]{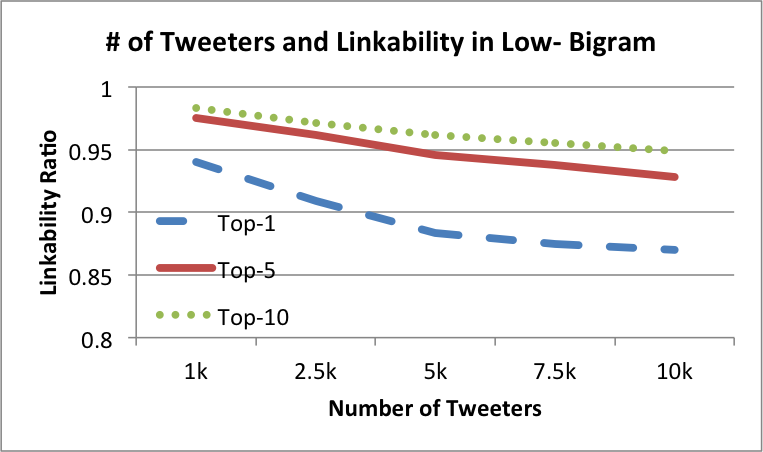}}
  \label{fig:bi-vary}
  \caption{LRs of bigram-based NB model  when varying the number of users in  $\Prol$ \subref{fig:ng-2-vary-prol} and $\Low$ \subref{fig:ng-2-vary-low}}
\end{figure}

\noindent
{\bf Bigrams.} With similar ranges, we vary the number of users  in  $\Prol$ and $\Low$. 
Figures \ref{fig:ng-2-vary-prol} and \ref{fig:ng-2-vary-low}  show LRs for different set sizes with AR size of $100$.  Interestingly, LR does not decrease much when we increase set size. For example, looking at the entire range, 
Top-1 LR does not decrease over 4\% and 7\% in $\Prol$ and $\Low$, respectively. Meanwhile, 
Top-1 LR stays above 95\% in $\Prol$ and above 87\% in $\Low$.  Also, in Top-5 and Top-10 LR, the decrease 
does not exceed 5\% in $\Low$ and almost 0-1\% in $\Prol$. This shows that the bigram-based NB model 
is very resilient against increasing the number of users. 
We believe that these results should be very troubling to tweeters worried about linkability of their tweets.

\subsection{Improving Unigram-based Model}
\label{sec:hashtag-and-n1}
Previously, we observed that relying on only bigram tokens yields very high LRs. 
However, bigrams also require more resources than unigrams: $676$ vs $26$ tokens.
Thus, bigram-based models are less scalable. To this end, we consider improving LRs 
when only unigrams are used, by exploring the use of hashtags. We first consider 
using unigrams from the hashtags themselves, and then combine them with unigrams 
from full-tweet texts.  \\

\noindent
{\bf Hashtags}
are a peculiar, yet popular, feature of Twitter. Not surprisingly, many tweets in our dataset contain one or more
hashtags.
We first filter out from $\Prol$ and $\Low$ all tweets that do not include any hashtags. We then discard all users with fewer than $300$ hashtag-containing tweets; this is so 
that we can populate their corresponding AR sets with 100 tweets\footnote{We defer to future work 
the case of linkability when IR size is smaller than AR size.}. This leaves us with 
$3,179$  and $160$ users in  $\Prol$ and $\Low$, respectively.  Since the resultant size of 
$\Low$ is small, we confine our analysis to $\Prol$. The number of tweets in filtered $\Prol$ is 
$4,274,188$.

Initially, we intended to use hashtags as tokens in the NB model. However, this resulted
in a very large number ($>150,000$) of hashtags, which is very resource-consuming. 
To remedy the situation, we decided to use unigram tokens within hashtags. 
Since hashtags can include $11$ non-alphabetical characters 
(i.e., $0-9$ and ``\_"), we ended up with $37$ tokens. 

\begin{figure}[thb!]
  \centering
  \includegraphics[height=2.0in,width=0.59\textwidth]{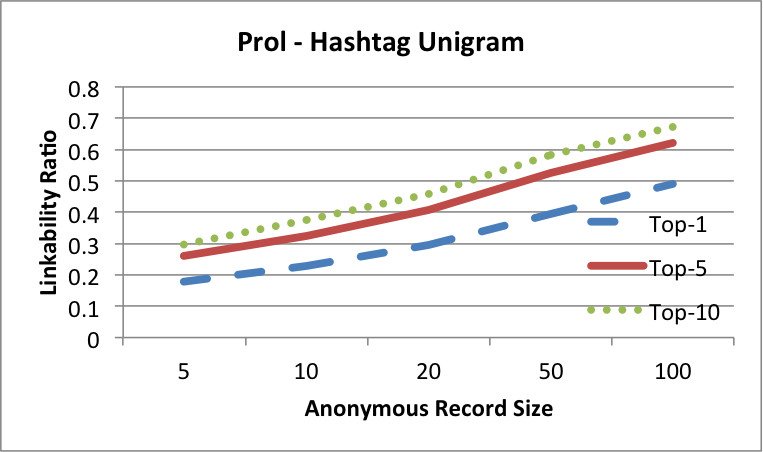}
  \caption{Top-1, Top-5, and Top-10 LRs of hashtag-based NB model when using unigrams }
  \label{fig:hash-n1}
\end{figure}

Figure \ref{fig:hash-n1} shows LRs for hashtag-based unigrams in NB\footnote{Recall that 
our analysis is based on the filtered version of $\Prol$ with $3,179$ users.}. 
As can be easily seen, Top 10 LR reaches 67\% for AR of $100$. Despite only relying on
hashtags, we can successfully link $2/3$ of ARs.  

\noindent
{\bf Combining Hashtags and Full-Tweet Texts.}

We consider improving LR by exploring unigrams of full-tweet texts
(a set of $26$) combined with hashtags-based unigram tokens -- a set of $37$.
Combining these two sets yields $63$ tokens, which is still a lot less than $676$ with bigrams.

As discussed in Section \ref{background}, we use the following $Log-Sum$ 
to sort the matching users from a set of tokens: 
\begin{figure}[thb!]
  \centering
  \includegraphics[height=2.0in,width=0.59\textwidth]{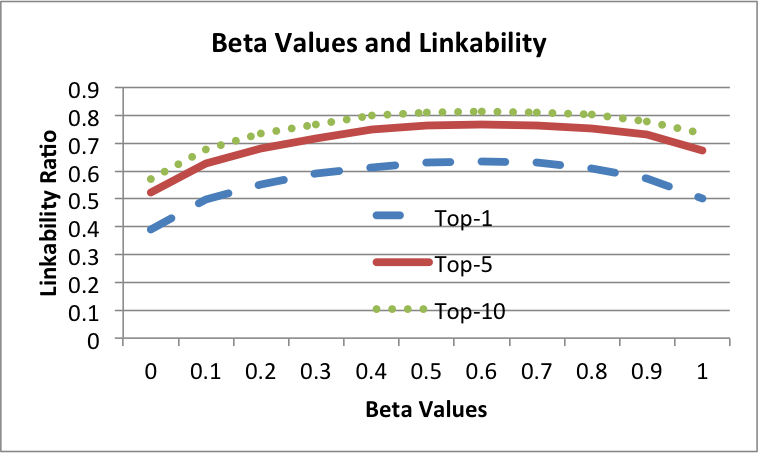}
  \caption{Top-1, Top-5 and Top-10 LRs  when varying beta $\beta$ values from 0 to 1}
  \label{fig:train-top}
\end{figure}
$$
\footnotesize 
\sum_i logP(t_i|U) \nonumber 
\normalsize
$$
We now use a weighted average for combining the
$Log-Sum$ of full-tweet-text-based tokens and hashtag-based tokens as follows:
\scriptsize
$$ 
(\beta)\times\sum_{t_i^{tweet}}logP(t_i^{tweet}|U)+(1-\beta)\times\sum_{t_i^{hashtag}} logP(t_i^{hashtag}|U)\nonumber 
$$
\normalsize
However, there is no clear way to assign a value to $\beta$. We 
experimented with several choices. Specifically, we tried all $\beta$ values 
ranging in $[0-1]$ at $0.1$ increments and observed  the highest LR with $\beta=0.6$ 
in Top-1, Top-5 and Top-10 LRs. Figure \ref{fig:train-top} shows LRs for different $\beta$ values.
Note that $\beta$ selection process (training) 
is restricted to the set of IRs (ARs are excluded). That is, $\beta$ is learned using
IRs, by further splitting them into identified and anonymized\footnote{We allocate the 
last 50 tweets of IRs as the new ARs used for training.} sets, and computing the 
corresponding LR values in the filtered version of $\Prol$.  

\begin{figure}[thb!]
  \centering
  \subfigure[]{\label{fig:combine-hash-top-1}\includegraphics[height=2.0in,width=0.32\textwidth]{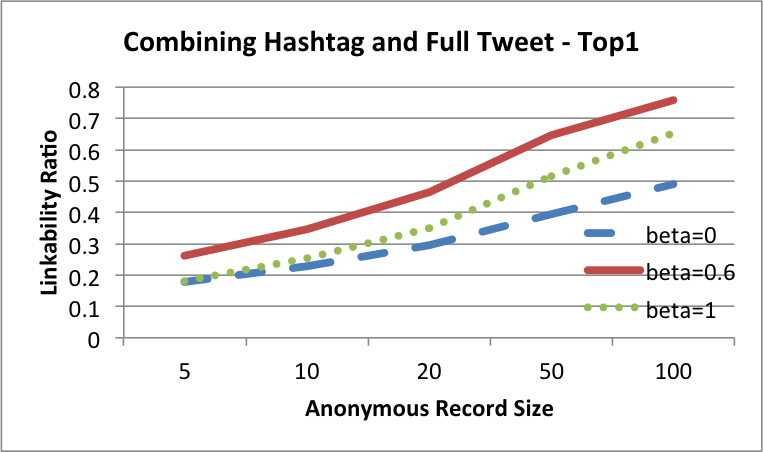}}
  \subfigure[]{\label{fig:combine-hash-top-5}\includegraphics[height=2.0in,width=0.32\textwidth]{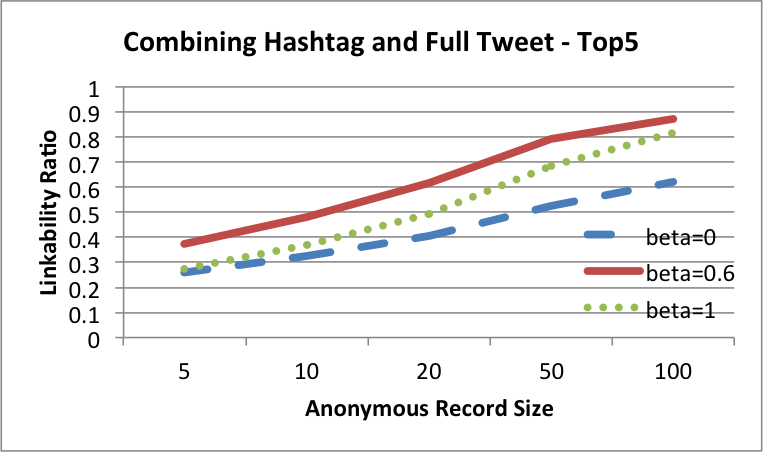}}
  \subfigure[]{\label{fig:combine-hash-top-10}\includegraphics[height=2.0in,width=0.32\textwidth]{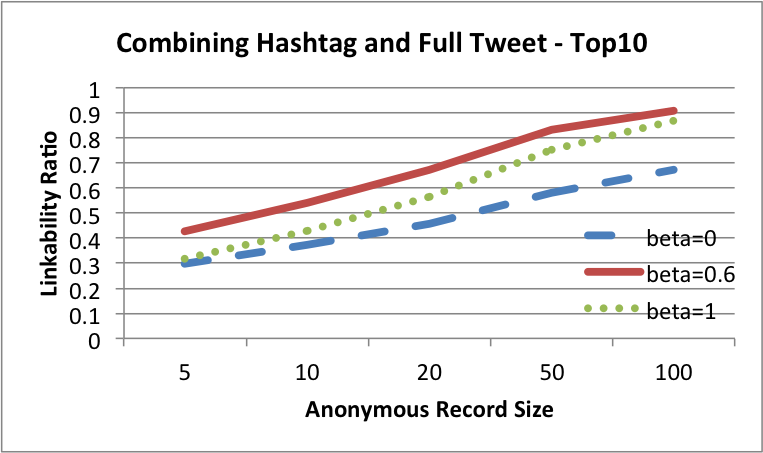}}
  \label{fig:hash-and-n1-82k}
  \caption{Top-1 \subref{fig:combine-hash-top-1} , Top-5 \subref{fig:combine-hash-top-5} and Top-10 \subref{fig:combine-hash-top-10} LRs of the revised version of $\Prol$ when combining unigrams of full-tweet texts and hashtags}
\end{figure}

\noindent
{\bf Results of Combining.} 
Having chosen $\beta=0.6$, we used it in computing the weighted average. 
Figures \ref{fig:combine-hash-top-1}, \ref{fig:combine-hash-top-5} and \ref{fig:combine-hash-top-10} 
show Top-1, Top-5 and Top-10 LRs, when combining unigrams of full-tweet texts and hashtags for 
the filtered $\Prol$. All figures show LR for $\beta$ set to $0$ (hashtags only), 
$1$ (full-tweet texts only), and $0.6$ (combination of tweet texts and hashtags).
As evident from the figures, combining full-tweet texts with hashtags substantially boosts 
LRs for  all AR sizes. The improvement over the best of two other curves ranges from 
$8-13\%$, $6-13\%$ and $4-11\%$ in Top-1, Top-5 and Top-10 LRs, respectively.  
Note that LRs, when $\beta=1$, are different from that in Section \ref{sec:uni}. That is because 
we are assessing the LRs
of filtered $\Prol$, which is much smaller (in terms of number of tweets for IRs) than $\Prol$. We conclude that unigrams in hashtags make tweets linkable, but they are more powerful when combined with unigrams from full-tweet texts. 
This combination definitely depends on the choice of $\beta$.   

\subsection{Considering Dual-Account Tweeters}
\label{realworld}
So far, we analyzed many sets of tweets, each set authored by a distinct user corresponding to a 
Twitter account. After {\em artificially} splitting each such set into Identification Record (IR) and 
Anonymous Record (AR), we discovered that they are highly linkable. 
In practice, our technique aims to link {\bf distinct users}, i.e., multiple bodies 
of tweets emanating from different Twitter accounts. Therefore, results discussed
above can be criticized for being too artificial. Indeed, if our approach is truly effective,
it must be evaluated using some ``ground truth'', i.e., real users who tweet via multiple 
accounts. In this section, we apply the NB classifier to exactly this type of data.

Clearly, we can not assemble a comprehensive collection of all multi-account bodies of tweets,
even for a fixed period of time. Instead, as ground truth data, we use a fairly small number of
account-pairs that are known to be operated by the same author/user. We merge (or mix in)
this new information into the much larger set of tweets used in the previous sections, and 
then re-apply our NB classifier.

\subsubsection{Multi-Account Dataset}
In order to collect tweets from dual-account authors, we manually (and extensively) searched the web
for public information pertaining to individuals who operate multiple accounts..
For example, we used search variations of keywords, such as ``multiple twitter accounts",
and found several blog posts where Twitter users publicly reveal tweeting via multiple accounts.
This netted us a total of $41$ accounts operated by $14$ distinct Twitter users. Of these, 
$11$ actively maintain $2$ accounts, while remaining $3$ users have $3$ or more accounts.
For each account, we collect corresponding sets of tweets via Twitter API with 
twitter-logger\footnote{https://dev.twitter.com/ and https://github.com/sixohsix/twitter}.

Because of quotas in Twitter API, we could only dump up to $\sim$$3200$ of a given account's 
most recent tweets. After crawling $41$ accounts, we observed an average number of tweets per
account of $1,631$, with a maximum of $3,241$ and a minimum of $34$.
For each multi-account user, we considered a set of most-prolific two accounts, that
have the highest number of tweets. This set constitutes $14$ dual-account owners, 
along with their corresponding tweets, referred as $\Dual$.
Most users in $\Dual$ operate one relatively general personal account and another that is 
more focused on a specific topic, e.g.,  professional, hobbies, political, or sports.

\subsubsection{Linkability Results}
Based on results discussed in the earlier part of this paper, we again use bigram NB 
as the linking model for the $\Dual$ dataset. As a sanity check, we first assessed NB's
performance with $\Dual$ without mixing in tweets from any external dataset. 
For each dual-account user, we randomly selected one of the two accounts (i.e, 
all tweets therein) as IR, and the other -- as AR.

\begin{figure}[thb!]
  \centering
  \includegraphics[height=2.0in,width=0.59\textwidth]{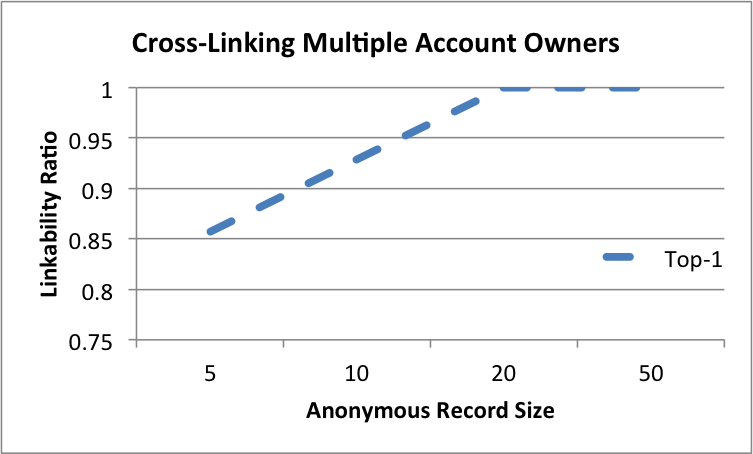}
  \caption{Top-1 LR in $\Dual$ -- Total number of dual-account owners is 14.}
  \label{fig:real-world-link-14}
\end{figure}

As before, we varied AR size between $5$ to $50$, by randomly selecting tweets\footnote{The 
maximum AR size was set to $50$, instead of earlier $100$, since a few accounts had
less than 100 tweets.}. Figure \ref{fig:real-world-link-14} shows LRs of $14$ users in $\Dual$. 
Top-1 LR is 100\% for AR size of at least $20$. Also, Top-1 LR exceeds 85\% for AR size less than $10$.
We therefore conclude that the NB bigram model is very successful in linking tweets from different accounts.

\begin{figure}[thb!] 
  \centering
  \subfigure[]{\label{fig:real-world-mix-pro}\includegraphics[height=2.0in,width=0.49\textwidth]{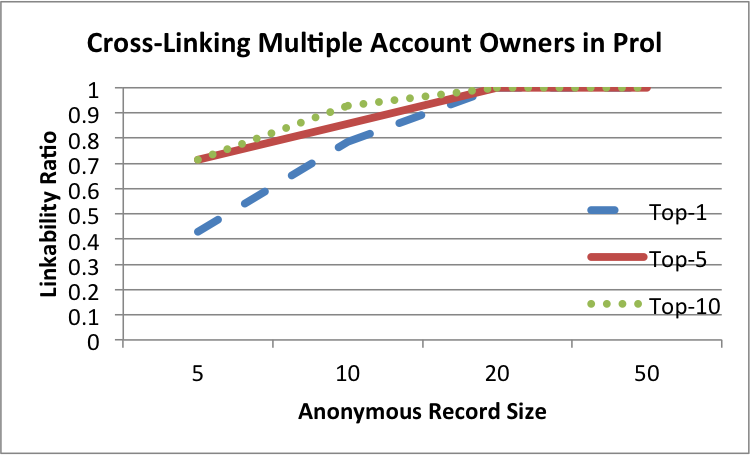}}
  \subfigure[]{\label{fig:real-world-mix-low}\includegraphics[height=2.0in,width=0.49\textwidth]{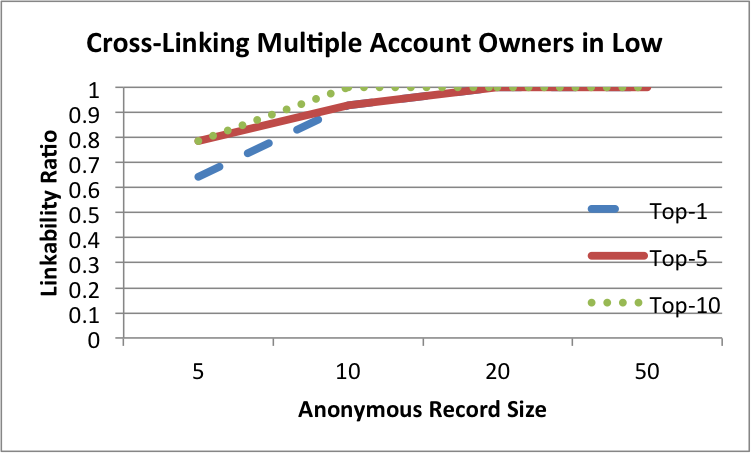}}
  \caption{Top-1, Top-5 and Top-10 LRs in $\Dual$ when merging accounts with $\Prol$ -- \subref{fig:real-world-mix-pro} and $\Low$ -- \subref{fig:real-world-mix-low}}
  \label{fig:real-world-mix}
\end{figure}

As the next step, we verify that the classifier is scalable, i.e., performs well
if tweets from $\Dual$ are merged with $\Prol$ and $\Low$ datasets, respectively. 
Specifically, we merge IRs from $\Dual$ and $\Prol$ / $\Low$.
Likewise, we augment ARs of $\Dual$ with ARs of $\Prol$ / $\Low$.
Figures \ref{fig:real-world-mix-pro} and \ref{fig:real-world-mix-low} show
LRs of 14 dual-account owners in $\Dual$ augmented by $\Prol$ and $\Low$, 
respectively\footnote{The total number of users is $8,262+14=8,276$ in 
Figure \ref{fig:real-world-mix-pro} and $10,000+14=10,014$ in \ref{fig:real-world-mix-low}.}.
Once again, Top-1, Top-5 and Top-10 are at 100\% when AR size exceeds 20.
Surprisingly, for AR size of $10$, Top-10 LR exceeds $90\%$ for both cases.
This clearly confirms effectiveness of our NB bigram model for the dual-account 
user linkage.

\section{Key Results}
\label{sec:summary}
Key elements of our analysis can be summarized as follows:
\begin{enumerate}
\item  Tweets are highly linkable. Top-1/Top-10 LRs reach up to 95/99\% for a large set of users 
(over $8,000$) by only relying on bigrams; see Section \ref{sec:di}.    
\item  Tweets remain highly linkable even in the context of only unigrams. Top-10 LR exceeds 97\% and 
85\% in $\Prol$ and $\Low$, respectively; see Section \ref{sec:uni}. 
\item ARs are highly linkable even when their sizes are small. For example, with bigrams and AR size of 
$20$, Top-10 LRs exceed 95\% and 79\% in  $\Prol$ and $\Low$; see Section \ref{sec:di}.  
\item  LRs are the highest in $\Prol$. However, even with less prolific tweeters, we obtain high LRs. 
For example, in $\Low$, Top-10 LR reaches 95\%, 96\% and 98\%, for the cases of:
$10,000$, $5,000$ and $1,000$ users, respectively (with bigrams); see Section \ref{sec:vary}. 
\item Again, with bigrams, LRs do not decrease beyond 7\% and 4\% in $\Prol$ and $\Low$, 
if we increase the number of users from $1,000$ to the full user set size;
see Section \ref{sec:vary}. 
We believe these results are troubling to privacy-conscious tweeters.  
\item Unigram-based LR can be substantially improved by combining the 
unigram model derived from full-tweet texts with that derived from hashtags;
see Section \ref{sec:hashtag-and-n1}.
\item We evaluated our approach with data from a small number of actual dual-account users.
The linkability ratio can reach 100\% for all $14$ dual-account owners
that we used as the ``ground truth" dataset; see Section \ref{realworld}.
\end{enumerate}

\section{Related Work}
\label{related}
\noindent
{\bf Author Attribution in Twitter.}
Some prior results focused on authorship identification and stylometric analysis of microblogging, e.g.,
\cite{twaznme,attribution-shortmsgs,twitter-attr-140-less,koppel-authorattrib,aswasa-authorattrib}. 
\cite{twaznme} considered re-identifying authorship of tweets from 
a set of three authors, while using over $5,000$ dimensions as input for a 
Support Vector Machine (SVM) classifier. Similarly, \cite{twitter-attr-140-less} 
investigated pseudonymity for a set of $50$ Twitter users. 
 \cite{attribution-shortmsgs} studied the use of n-grams with Na\"ive Bayesian as the linkability model. 
In particular, 2- to 6-grams are evaluated and a $98\%$ linkability is achieved in the setting of $50$ authors. 
An identification technique based on extracting a set of lexical and syntactical features along with SVM 
is proposed in \cite{aswasa-authorattrib}. It achieves $91\%$ accuracy for a set of $15$ authors. 
%
%
Moreover, a technique based on extracting a set of textual signatures  per author is proposed in \cite{koppel-authorattrib}. Character n-gram 4 and word n-grams are used as features and an accuracy of 30 is achieved  when the number of authors is 1,000 (70\% accuracy when the number of authors is 50). 

There are four main differences between aforementioned results and our work First, 
we assess linkability on a large scale. The numbers of tweeters is way larger 
 than in prior related studies. Second, we use unigrams, which drastically reduces the 
number of tokens. Third, we also include hashtags and show their effectiveness in linkability 
when used alone or in combination with other unigrams. Finally, we successfully 
re-produce and confirm linkability results for a small set of actual dual-account twitter users.   

\noindent
{\bf Cross-Linking Accounts.}
\cite{Goga-link} reports on efforts to cross-link accounts between different social networks, in particular,
from Yelp to Twitter and from Flickr to Twitter. Geo-location information, timestamps and text-based features 
are used in linking user accounts, along with the cosine distance function. There are some notable differences 
between this study and ours. 
Techniques that link accounts across social networks might not work for linking accounts 
within the same network; we do the latter. 
%
%
One reason is that accuracy (across social networks) tends to be high when linking models use 
similarities among user-names and location information as features (Otherwise, it is quite low).
Whereas,  a user operating two separate accounts within the same social network would likely
not pick similar user-names. Also, location services might be disabled in some social networks. 

Other prior work that explored linkability of accounts across platform includes 
\cite{dperitoPETS11} and \cite{footprint}. 

\noindent
{\bf Author Attribution.}
A recent effort \cite{herbert-deanonymizer} investigated de-anonymization of ``anonymous''
peer reviews of academic papers. The best result achieved is close to $90\%$. 

One of the best-known results in this area is \cite{writeprints-abbasi} which uses 
an extensive set of features, called Writeprints. This approach
attained identification accuracy of $91\%$. Other related studies include \cite{mish12} 
which explored author linkability in user reviews using similar (to ours) probabilistic 
techniques.  \cite{internet-scale} tackled large-scale online authorship re-identification 
using linguistic stylometry techniques, achieving up to $80\%$ accuracy.  
We refer to \cite{author-survey} for an extensive survey of author
identification and authorship attribution literature.

\section{Discussion}
\label{sec:disc}
%
%
Although there are many stylometric features in the literature, the focus of this paper is 
on {\em scalable} linkability analysis, which motivated us to consider only simple features,
allowing the models to perform well with the large number of users. 

Even though our technique was evaluated over a very large dataset with numerous accounts, 
it is possible that some accounts we ``linked'' are actually operated by a group of authors, 
rather than by a single one. This might occur if an account serves as an outlet for an organization, 
e.g., a business, a government agency or a professional society. While privacy concerns are 
clearly much less serious with such accounts, we believe it is very helpful for their owners 
to be aware of linkability issues. 

Finally, recall that our analysis considered the maximum AR size of 100. While this 
might seem large, it represents $\leq 5\%\;$ $(\leq 30\%)$ of the total tweets of the user 
with the minimum contribution in \Prol $\;$ (\Low). In practice, it is easy to collect a 
daily global snapshot of all tweets by using the dedicated Twitter 
API\footnote{Accessible via: \url{https://dev.twitter.com/docs/api/1.1/get/search/tweets}}. 
%

\section{Future Work}
\label{future-work}
%
As part of future work, we plan to improve
LRs for smaller AR sizes by considering probabilistic models other than NB
(with dependencies among some features) and by looking into other features,
such as ratio of tweets to retweets and corse-grained categories of hashtags. We 
also would like to perform linkability analysis on different
author demographics, such as those with very few tweets.
Moreover, we believe that further analysis is needed to understand what makes tweets more (or less)
linkable. This would allow us to make concrete recommendations for
tweeters to retain more privacy.



\bibliographystyle{IEEEtran}
\bibliography{IEEEabrv,paper}
\end{document}